\documentclass[conference,9pt]{IEEEtran}

\usepackage[utf8]{inputenc}
\usepackage{booktabs} 
\usepackage{graphicx}
\usepackage{subfigure}
\usepackage{units}
\usepackage{blindtext}
\usepackage{xcolor,xspace}
\usepackage{amsmath}
\usepackage{paralist}
\usepackage{amsmath}
\usepackage{tikz}
\usepackage{booktabs}
\usepackage{pifont}
\usepackage{multirow}
\usepackage{hhline}
\usepackage{balance}
\usepackage{footnote}
\usepackage{logicproof}
\usepackage{amsthm}
\usetikzlibrary{patterns}
\usetikzlibrary{calc}
\usetikzlibrary{decorations.pathreplacing}
\usepackage{pbox}

\usepackage{arydshln}
\usepackage{mdframed}
\usepackage{pgf-umlsd}

\usepackage[T1]{fontenc}
\usepackage{
  beramono,
  listings,
  textcomp
}

\usetikzlibrary{arrows,automata}
\usetikzlibrary{calc}
\usetikzlibrary{arrows,calc,shapes,decorations.pathreplacing}

\usepackage{listings}
\usepackage[ruled]{algorithm2e}
\usepackage{algorithmic}

\LinesNumbered

\usepackage{url}

\expandafter\def\csname ver@l3regex.sty\endcsname{}
\usepackage[
    n,
    advantage,
    operators,
    sets,
    adversary,
    landau,
    probability,
    notions,
    logic,
    ff,
    mm,
    primitives,
    events,
    complexity,
    asymptotics,
    keys]{cryptocode}

  \newcolumntype{P}[1]{>{\centering\arraybackslash}p{#1}}

\newcommand{\acro}{{{\sf\it DIALED}}\xspace}
\newcommand{\tinycfa}{{{\sf\it Tiny-CFA}}\xspace}

\newcommand{\prv}{{\ensuremath{\sf{\mathcal Prv}}}\xspace}
\newcommand{\vrf}{{\ensuremath{\sf{\mathcal Vrf}}}\xspace}

\newcommand{\RA}{{\ensuremath{\sf{\mathcal RA}}}\xspace}
\newcommand{\CFA}{{\ensuremath{\sf{\mathcal CFA}}}\xspace}
\newcommand{\DFA}{{\ensuremath{\sf{\mathcal DFA}}}\xspace}

\newcommand{\rtop}{\ensuremath{\mathcal R}\xspace}

\newcommand{\program}{\ensuremath{\mathcal Op}\xspace}

\newcommand{\pox}{\texttt{PoX}\xspace}

\mathchardef\mhyphen="2D

\newcommand{\cflog}{{\small CF-Log}\xspace}
\newcommand{\ilog}{{\small I-Log}\xspace}

\usepackage{tikz}
\def\checkmark{\tikz\fill[scale=0.4](0,.35) -- (.25,0) -- (1,.7) -- (.25,.15) -- cycle;} 

\newtheorem{definition}{Definition}

\begin{document}

\title{\acro: Data Integrity Attestation for \\ Low-end Embedded Devices}

\author{\IEEEauthorblockN{Ivan De Oliveira Nunes}
\IEEEauthorblockA{\textit{University of California, Irvine} \\
ivanoliv@uci.edu}
\and
\IEEEauthorblockN{Sashidhar Jakkamsetti}
\IEEEauthorblockA{\textit{University of California, Irvine} \\
sjakkams@uci.edu}
\and
\IEEEauthorblockN{Gene Tsudik}
\IEEEauthorblockA{\textit{University of California, Irvine} \\
gene.tsudik@uci.edu}
}

%
\maketitle
%

%
\begin{abstract}
Verifying integrity of software execution in low-end micro-controller units (MCUs) is a well-known open problem. 
The central challenge is how to securely detect software exploits with minimal overhead, since these MCUs are designed 
for low cost, low energy and small size. Some recent work yielded inexpensive hardware/software co-designs
for remotely verifying code and execution integrity. In particular, a means of detecting unauthorized code modifications and control-flow 
attacks were proposed, referred to as Remote Attestation (\RA) and Control-Flow Attestation (\CFA), respectively.
Despite this progress, detection of {\bf data-only attacks} remains elusive. Such attacks exploit software vulnerabilities 
to corrupt intermediate computation results stored in data memory, changing neither the program code nor its control flow.
Motivated by lack of any current techniques (for low-end MCUs) that detect these attacks, in this paper we propose, implement
and evaluate \acro \footnote{\textbf{To appear at DAC'21}}, the first Data-Flow Attestation (\DFA) technique applicable to the most resource-constrained embedded 
devices (e.g., TI MSP430). \acro works in tandem with a companion \CFA scheme to detect all (currently known) types of runtime 
software exploits at fairly low cost.
\end{abstract}

\section{Introduction}\label{sec:intro}
Embedded systems are growing in number and variety in recent years, rapidly becoming ubiquitous in
many aspects of modern society. 
Some are ultra-cheap and specialized, built atop low-energy, low-cost, and tiny MCUs,
e.g., TI MSP430 and AVR ATMega32. Despite being extremely resource-constrained, these MCUs
often perform safety-critical tasks that involve sensing and/or actuation. Hence, their security is very important.

Securing low-end MCUs is very challenging, since they often lack any security-relevant hardware features.
At the lowest end of the spectrum, devices have no MMUs or MPUs, and thus have no means to support an 
OS or even a microkernel. Hence, sophisticated malware prevention and detection techniques that work on smart-phones, 
laptops, desktops, and servers are inapplicable.  To address this problem, a number of
architectures~\cite{smart,vrasedp,apex,simple,KeJa03,SPD+04,SLS+05,SLP08,tiny-cfa} were proposed to
support rudimentary security services on low-end MCUs. One prominent research direction involves so-called
hybrid (hardware/software co-design) architectures that offer strong security guarantees at minimal hardware cost.

In that vein, architectures for Remote Attestation (\RA)~\cite{smart,vrasedp,simple}, Proofs of Execution (\pox)~\cite{apex}, 
and Control-Flow Attestation~\cite{tiny-cfa} are of particular interest. \RA is an interaction between a trusted and more powerful entity,
called a Verifier (\vrf) and a potentially compromised remote low-end device, called a Prover (\prv). It allows \vrf to 
securely measure \prv's memory contents, enabling detection of malware that modifies code installed on \prv. 
\pox augments \RA by providing \vrf with a proof that the attested code was executed and that any claimed outputs (e.g., sensed 
values) were indeed produced by executing the expected code. Finally, \CFA detects control-flow attacks, whereby  
benign code on \prv contains unknown bugs (e.g., buffer overflows caused by lack of array bound checks) that can be 
exploited to hijack the program's control-flow, i.e., the order of instruction execution. In summary, \CFA provides \vrf with a 
proof that benign code was executed in a particular valid or expected order.

Despite these advances, detection of data-only attacks remains elusive.
Similar to control-flow attacks, data-only attacks exploit vulnerabilities in benign code to corrupt intermediate values in 
data-memory. It is well-known~\cite{runtime_attacks_sok,ispoglou2018block} that data-only attacks need not alter the code or its control-flow in order to corrupt data. 
Without a way to detect data-only attacks (as well as code and control-flow modifications) 
results of \prv's remote computation cannot be trusted.

Very recently, a novel technique supporting both \CFA and \DFA, called OAT~\cite{oat}, was proposed. However, it 
implements \DFA by relying on trusted hardware support from ARM TrustZone, which is only available on higher-end 
platforms (e.g., smartphones, Raspberry Pi, and similar) and is not affordable to low-end, low-energy MCUs.
In addition, OAT's security relies on the application programmer's ability to correctly annotate all critical variables 
in the code to be attested. This is a strong assumption, since most control-flow and data-only exploits are caused by 
implementation bugs introduced by the very same application programmer. Naturally, it would be beneficial for this assumption
to be avoided.

In this paper, we focus on security against data-only attacks on low-end MCUs and propose \acro:  
\underline{D}ata \underline{I}ntegrity in \underline{A}ttestation 
for \underline{L}ow-end \underline{E}mbedded \underline{D}evices.
\acro's only hardware requirement is that already provided (at relatively low-cost) by the \pox architecture APEX~\cite{apex}. 
\acro uses APEX to securely log and authenticate any data inputs used by the program.
This authenticated log allows \vrf to reconstruct the entire data-flow of the program's execution, thus
enabling detection of any data-corruption attacks via abstract execution of the attested program.

Similar to OAT, \acro is implemented alongside Tiny-CFA~\cite{tiny-cfa}, a low-cost \CFA technique. 
This composition enables, for the first time, detection of both control-flow and data-only attacks for low-end MCUs.
Notably, \acro does not rely on code annotation; thus, its security neither requires, nor depends on, any human intervention.
In the rest of this paper, we describe \acro's design and analyze its security. We also report on the implementation of 
\acro along with Tiny-CFA on the TI MSP430 MCU and demonstrate its cost-effectiveness by
providing both \CFA and \DFA for three applications.

\section{Background \& Problem Statement}\label{sec:problem_statement}
This section overviews targeted devices and defines the problem setting.
\subsection{Scope of Low-End MCUs}\label{sec:scope}
This paper focuses on tiny CPS/IoT sensors and actuators, or hybrids thereof.
These are some of the smallest and weakest devices based on low-power single-core MCUs with small 
program and data memory (e.g., aforementioned Atmel AVR ATMega and TI MSP430), with
$8$- and $16$-bit CPUs running at $1$-$16$MHz, with $\approx64$ KBytes of addressable memory.
SRAM is used as data memory, normally ranging between $4$ and $16$KBytes, while the rest of the
address space is available for program memory.  Such devices usually run software atop ``bare metal'', 
execute instructions in place  (physically from program memory), and lack any memory management unit (MMU) 
to support virtual memory.
%

\subsection{Control-Flow vs. Data-Only Attacks}\label{sec:background_attacks}
Both control-flow and data-only attacks violate program execution integrity without modifying the actual executable, by
taking advantage of implementation bugs, e.g., lack of array bound checks. Such vulnerabilities are quite 
common in memory-unsafe languages, such as \texttt{C}, \texttt{C++}, and Assembly, which are widely used to 
program MCUs.

\lstset{language=C,
	basicstyle={\tiny\ttfamily},
	showstringspaces=false,
	frame=single,
	xleftmargin=2em,
	framexleftmargin=3em,
	numbers=left, 
	numberstyle=\tiny,
	commentstyle={\tiny\itshape},
	keywordstyle={\tiny\ttfamily\bfseries},
	keywordstyle=\color{purple}\tiny\ttfamily\ttfamily,
	stringstyle=\color{red}\tiny\ttfamily,
        commentstyle=\color{black}\tiny\ttfamily,
        morecomment=[l][\color{magenta}]{\%},
        breaklines=true
}
\begin{figure}[hbtp]
\centering
\begin{minipage}{0.7\linewidth}
\begin{lstlisting}[]
int dose = 0;

void injectMedicine(){
  if (dose < 10){  //safety check preventing overdose
    P3OUT = 0X1;	
    delay(dose*time_per_dose_unit);
  }
  P3OUT = 0x0;
}

void parseCommands(int *recv_commands, int length){
  int copy_of_commands[5];
  memcpy(copy_of_commands, recv_commands, length);
  dose = processCommands(copy_of_commands);
  return;
}
\end{lstlisting}
\end{minipage}
\caption{Embedded application vulnerable to a control-flow attack~\cite{tiny-cfa}.}\label{fig:cf_attack}
\end{figure}

Control-flow attacks change the order of instructions execution thus changing program behavior, escalating 
privilege, and/or bypassing safety checks. Consider the example in Figure~\ref{fig:cf_attack} (taken from~\cite{tiny-cfa}).
In this embedded application, the MCU is connected through the general purpose input/output (GPIO) port $P3OUT$ 
(lines $5$ and $8$) to an actuator that injects a certain dose of medicine, determined in software, according to 
commands received from the network, e.g., from a remote physician.

The function \texttt{injectMedicine} injects appropriate dosage given by the variable \texttt{dose}, by triggering 
actuation for period of time proportional to the value of \texttt{dose}.
To guarantee a safe dosage, the \texttt{if} statement (line 4) assures that the maximum injected dosage is $9$, 
thus preventing the patient from over-dosing due to errors.
The function \texttt{parseCommands} (line 11) makes a copy of received commands and processes them to determine 
the appropriate dosage. However, this function can be abused by a control-flow attack at line 13.
Specifically, because \texttt{copy\_of\_commands} has a fixed length of $5$, an input array of size more than $5$
causes a buffer overflow corrupting the data in stack, including the return address of \texttt{parseCommands}. 
In particular, the return address can be overwritten with the value of \texttt{recv\_commands[5]}. 
By setting the content of \texttt{parseCommands[5]} to be the address 
of line 5, such an attack causes the control flow to jump directly to line 5, skipping the safety 
check at line 4, and potentially overdosing the patient.

As mentioned earlier, \CFA securely logs all control-flow deviations  and (upon request) provides an authenticated copy of this log
to \vrf, enabling detection of the aforementioned attack. We overview a concrete \CFA architecture in 
Section~\ref{sec:background_cfa}. However, \CFA cannot detect data-only attacks that do not change the control-flow.
Figure~\ref{fig:df_attack} presents a second implementation attempt, vulnerable to data-only attacks.
To see this vulnerability, note that P3OUT register controls multiple physical ports, each associated with one bit in 
P3OUT register, e.g.:
\begin{compactitem} {\footnotesize
 \item Setting $\mathsf{P3OUT = 00000000 = 0x0}$ turns all physical ports off
 \item Setting $\mathsf{P3OUT = 00000001 = 0x1}$ turns on 1st physical port 
 \item Setting $\mathsf{P3OUT = 00000010 = 0x2}$ turns on 2nd physical port 
 \item Setting $\mathsf{P3OUT = 00000011 = 0x3}$ turns on both 1st and 2nd physical ports, etc.
 }
\end{compactitem}
In order to trigger actuation through the proper port (Port 1 in this example), this code needs to set $\mathsf{P3OUT=0x1}$. 
This is configured in the global variable \texttt{set}, at line 1 of Figure~\ref{fig:df_attack}. The value of \texttt{set} is later used to 
trigger actuation at line 8. The code also allows \texttt{settings} to be updated at an arbitrary position defined by 
the input parameter \texttt{index}.
Since \texttt{settings} has a fixed length of $8$, a malicious input with $\texttt{index}=8$ would overflow this buffer 
causing \texttt{set} to be overwritten with the value of \texttt{new\_setting}. An input $\texttt{new\_setting=0}$ with 
$\texttt{index}=9$ would overwrite $\texttt{set}=0$. Later, at line 8, when $set$ is used to trigger actuation of port 1, 
it will instead have no actuation effect (since it is now $0$). Consequently, the medicine will not be injected.
It is important to note that this attack does not change the program control flow, but just corrupts data. It therefore can not 
be detected by \CFA alone.

\lstset{language=C,
	basicstyle={\tiny\ttfamily},
	showstringspaces=false,
	frame=single,
	xleftmargin=2em,
	framexleftmargin=3em,
	numbers=left, 
	numberstyle=\tiny,
	commentstyle={\tiny\itshape},
	keywordstyle={\tiny\ttfamily\bfseries},
	keywordstyle=\color{purple}\tiny\ttfamily\ttfamily,
	stringstyle=\color{red}\tiny\ttfamily,
        commentstyle=\color{black}\tiny\ttfamily,
        morecomment=[l][\color{magenta}]{\%},
        breaklines=true
}
\begin{figure}[hbtp]
\centering
\begin{minipage}{0.78\linewidth}
\begin{lstlisting}[]
int set = 0x1; // configured to cause actuation on Port 1
int settings[8]; // default settings produce dose = 5ul

void injectMedicinePort1(int new_setting, int index){
  settings[index] = new_setting;
  int dose = defineDosage(settings);
  if (dose < 10){  //safety check preventing overdose
    P3OUT = set;	
    delay(dose*time_per_dose_unit);
  }
  P3OUT = 0x0;
  return;
}
\end{lstlisting}
\end{minipage}
\caption{Embedded application vulnerable to a data-flow attack.}\label{fig:df_attack}
\vspace{-0.1cm}
\end{figure}

\begin{figure*}
	\begin{center}
	\hspace*{0.4cm}\includegraphics[width=1.6\columnwidth]{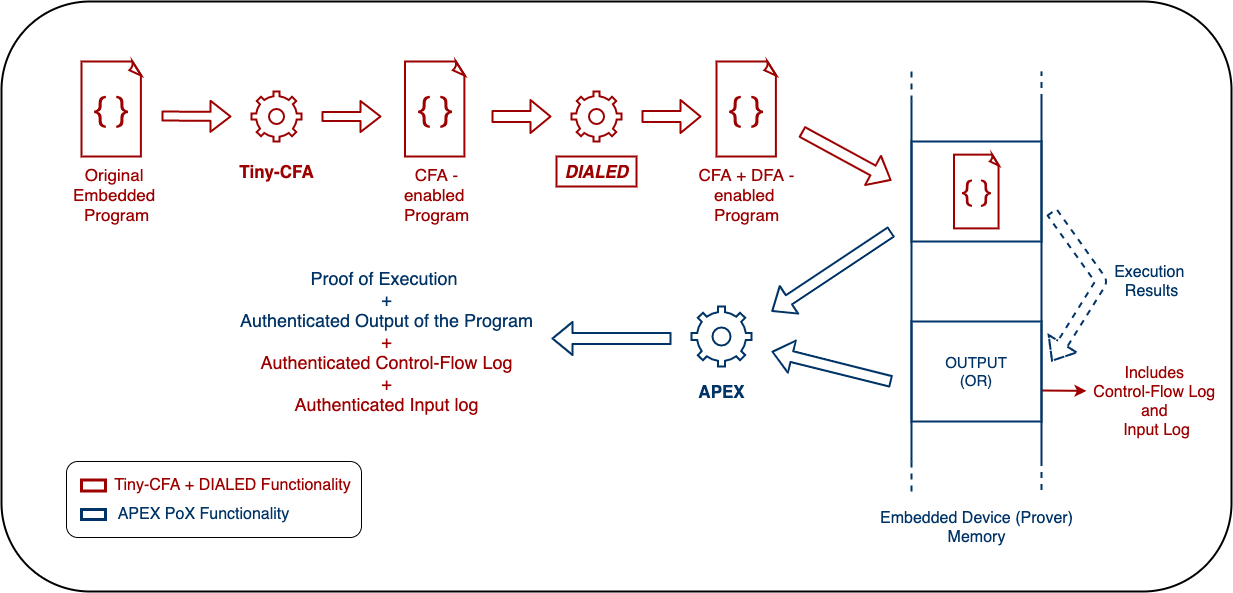}
	\end{center}
	\vspace{-1em}
	\caption{\acro Architectural Components.}\label{fig:components}
\end{figure*}

\subsection{Prior Work: \CFA}\label{sec:background_cfa}
In recent years several \CFA techniques have been proposed ~\cite{cflat,dessouky2017fat,dessouky2018litehax,zeitouni2017atrium,oat}. 
We focus on \tinycfa, the only \CFA for low-end devices targeted in this work.
\tinycfa is constructed atop another recently proposed technique --  APEX formally verified architecture~\cite{apex} that 
implements the so-called Proof-of-Execution (\pox) primitive. \tinycfa imposes no additional
hardware requirements, beyond those imposed by APEX. The \pox primitive allows \prv to
prove to \vrf cryptographically binding the following guarantees:
\begin{compactenum}
 \item The proper/expected code is loaded at a particular location in \prv's program memory, referred to as 
 Executable Range (ER).
 \item This proper/expected code was executed in its entirety: from its first instruction = its legal entry point, 
 until its last instruction = its legal exit.
 \item All outputs are authentic and were indeed produced by this execution. The outputs are written to a specific memory 
 location, referred to as Output Range (OR).
\end{compactenum}
Sizes and locations of ER and OR are configurable, which enables \pox for arbitrary code and arbitrary outputs. 
APEX is formally verified to guarantee secure \pox even in case of a full software compromise of the underlying MCU.
APEX also prevents any external data memory modifications that could attempt to tamper 
with ER's state while the code in ER is running (e.g., via interrupts or DMA).

\tinycfa implements \CFA atop APEX by instrumenting the executable with instructions that log the control flow to APEX-authenticated 
OR region. Specifically, given an executable, \tinycfa instruments all control flow-altering instructions (e.g., jumps, branches, 
function calls, and returns) by logging the destination address of each such instruction to the OR region.
\tinycfa also assures that the log in OR cannot be modified.
This way, given APEX formal guarantees, \vrf securely learns the exact control flow of program execution on a remote \prv.

\acro also uses code instrumentation and augments \tinycfa with the detection of data-only attacks.
\acro's core idea is to log (and send to \vrf) all the inputs of the program during execution along with its control flow, which enables \vrf to emulate 
the entire execution of the attested program, allowing it to detect both control-flow and data-flow
attacks. \acro is detailed in the next section.

\section{\acro Design}\label{sec:design}
Figure~\ref{fig:components} shows the components of \acro: it is implemented alongside \tinycfa (itself based on 
APEX)  to provide both \CFA and \DFA. The executable is separately instrumented by both \tinycfa and \acro. 
APEX provides a proof of the execution of the \emph{instrumented executable}, serving as an authenticator for its output: 
a log containing the executable's control flow and its data inputs. 

\subsection{Overview}
To detect data-only attacks, \acro uses a novel input detection method via secure instrumentation of the executable. 
This instrumentation guarantees that all relevant data is logged during program execution; this is in addition to 
the control-flow log produced by \tinycfa. The underlying \pox primitive provides \vrf with a proof that this output 
(containing all relevant data and the control-flow path) was indeed produced by the execution of the expected 
(instrumented) code. In doing so, \acro provides \vrf with all information needed to abstractly execute this program 
locally and detect any code, control flow, or data compromises. The core feature of \acro is detection and secure logging
of every external input received during program execution, including input from peripherals, the network, GPIO, as well as 
data fetches from memory locations outside the executable's own state.

Similar to OAT~\cite{oat}, the goal is to attest \textbf{embedded operations}, i.e., finite and self-contained safety-critical functions 
called by the program's main loop. Examples include sensing and actuation tasks triggered by 
commands received through the network, as in Section~\ref{sec:background_attacks}.
Since embedded operations typically have well defined and reasonably small number of data inputs, \acro can efficiently save 
all inputs to an append-only log -- Input Log (\ilog).
\acro instrumentation assures that all {\em data inputs} are appended to \ilog. 
%

\begin{definition}[Data Inputs]\label{def:data_inputs}
Any value read from any memory location \underline{\textbf{outside}} of the attested program's current stack. 
The program's current stack is the region located within the current stack pointer value (top of the stack) and
the value of the stack pointer when the attested program was first called (based of the program's stack). 
It includes all local variables.
\end{definition}

According to this definition, read instructions that move/copy data from peripherals, network or GPIO
are considered as Data Inputs and written to \ilog, since these involve reads from memory outside of the program's stack. 
But reads that occur during regular computation, e.g., instructions which compute on local variables are not written to \ilog, as they are not inputs.
This approach makes the size of \ilog relatively small, which is confirmed by the real-world embedded operations considered in our evaluation in Section~\ref{sec:eval}.


Recall that \tinycfa instruments the executable to produce a Control-Flow Log (\cflog).
In \acro, both \cflog and \ilog are written to APEX-designated output region $OR$. 
Hence, \vrf is assured of the integrity of these logs. In addition, given the attestation guarantee, 
\vrf is also assured that the correct/expected instrumented code was executed to produce this log.
By knowing the code, its control-flow, and all inputs, \vrf can locally emulate its execution 
and verify all steps in this computation, as well as detect all data-only and control-flow attacks.


\subsection{Adversary Model}\label{sec:design}
We assume an adversary that controls \prv's entire software state, including code and data. 
It can modify any writable memory and read any memory that is not explicitly protected by 
hardware-enforced access controls, e.g., APEX rules. Program memory modifications can 
change instructions, while data memory modifications can trigger control-flow and data-only 
attacks arbitrarily. Adversarial modification attempts are allowed before, during, or after the execution.

\subsection{Design Rationale}\label{sec:design}
\acro's security is based on five features: \textbf{F1-F5}.
We describe them at a high level in this section and discuss how to realize an instance of 
\acro on MSP430 via automated code instrumentation in Section~\ref{sec:implementation}.

\begin{center}
\vspace{0.5mm}
\noindent\textbf{(F1) Integrity Proofs for Code, Instrumentation, and Output}
\vspace{0.5mm} 
\end{center}

As an instrumentation-based technique, \acro is only secure if any modifications to the instrumented code 
itself (e.g., removing instrumented instructions) is detectable. Detection of code modifications is already offered by 
the underlying APEX \pox architecture (see Section~\ref{sec:background_cfa}). APEX guarantees that every code 
modification is detected by \vrf. It also guarantees that any modification of the attested executable's output region $OR$ 
(which, in our case, includes \cflog and \ilog) can only be done by the attested executable itself, during its execution.

\begin{center}
\vspace{0.5mm}
\noindent\textbf{(F2) Integrity Proof for the Control Flow}
\vspace{0.5mm} 
\end{center}

Since \acro relies on instrumented instructions, these instructions can not be skipped, e.g., via control-flow violations.
Therefore, \tinycfa ensures that the control flow is logged to \cflog and whatever is written to \cflog can not be modified; 
see~\cite{tiny-cfa} for details. Hence, all attempts to skip the logging of any data inputs are detectable by \vrf using \cflog.
The integrity of \cflog itself is important to \acro's overall functionality, since \vrf needs both \cflog and \ilog in order to 
abstractly execute the program and verify the integrity of the execution.

\begin{center}
\vspace{0.5mm}
\noindent\textbf{(F3) Secure Logging of Data Inputs from Operation Arguments}
\vspace{0.5mm} 
\end{center}

To enable abstract execution by \vrf, any arguments passed to the program at invocation must be securely logged to 
\ilog. \acro automatically instruments the executable with Assembly instructions that copy all program arguments to \ilog.

\begin{center}
\vspace{0.5mm}
\noindent\textbf{(F4) Secure Logging of Runtime Data Inputs}
\vspace{0.5mm} 
\end{center}

In addition to arguments, data inputs can be obtained at runtime, e.g., sensed values read from GPIO, or packets
arriving from the network. Such inputs are received through peripheral memory, at a particular set of physical addresses 
in data-memory. To detect and log these inputs \acro instruments every \texttt{read} instruction to check whether the read
address is outside the program's stack.  
The range of the stack is determined by $[ls, hs]$, where $ls$ is the value of the stack pointer saved at the moment when 
execution starts (before the allocation of local variables), and $hs$ always reflects the current stack pointer, i.e., the top of the 
stack.

\begin{center}
\vspace{0.5mm}
\noindent\textbf{(F5) \ilog and \cflog Integrity}
\vspace{0.5mm} 
\end{center}

%
To ensure integrity of \cflog and \ilog, \acro must guarantee that control flow and data-only attacks do 
not overwrite these logs. Thus, we realize \ilog and \cflog as a single stack data structure inside 
OR, from the highest value ($OR_{max}$) growing downwards.
The pointer to the top of this stack is stored in a dedicated register \rtop.
Each instruction that alters the control flow or involves data input is instrumented (with additional instructions)
to push the relevant values (either control-flow destination or data input) onto the stack, i.e.:
%
\begin{compactenum}
 \item Write the value (destination of address or data input) to the location pointed by \rtop; and
 \item Decrement \rtop.
\end{compactenum}
At instrumentation time, assembly code is inspected to ensure that no other instructions use \rtop. 
In all practical code examples we inspected, executables have at least one free register available. 
If no such register exists, the code can be recompiled to free up one register.
Whenever a write operation occurs, it is checked for safety, by seeing if the address of the write is within 
the range $[\rtop,OR_{max}]$, i.e., the current range for \ilog and \cflog. If an illegal write occurs, 
execution is aborted and \vrf treats it as an attack. Since these ``write checks'' are already 
needed, and implemented, by \tinycfa, they can be used ``as is'' by \acro, at no additional instrumentation cost. 

\subsection{Security Analysis}\label{sec:security}
Let \program denote an embedded operation for which control-flow and data-flow need to be attested.
Feature {\bf F1} assures to \vrf that \program indeed executed, and that neither its executable (including instructions 
added by \acro's instrumentation) nor the output (OR) produced by this execution has been tampered with.
Feature {\bf F2} assures that all changes to the control-flow of \program are written to $OR$ at runtime.
Similarly, {\bf F3 \& F4} guarantee that any data inputs are also logged to $OR$.
Therefore, what we need to show is that, once written, control-flow and data input values in $OR$ can not be 
modified during the rest of \program execution. This is exactly the guarantee offered by {\bf F5}. 
Therefore, \acro features {\bf F1-F5} suffice to guarantee the integrity of $OR$ and \program's executable (stored in $ER$), 
including \ilog and \cflog, even in the presence of potential control-flow and data-only attacks.
Given the integrity of received \ilog and \cflog, \vrf can emulate execution of \program locally 
and reproduce any type of runtime attack 
that may have occurred during \program's actual execution in \prv.

\section{\acro Implementation}\label{sec:implementation}
As described in Section~\ref{sec:design}, features {\bf F1, F2} and {\bf F5} are provided by APEX and \tinycfa. 
Hence, we focus on the implementation of {\bf F3-F4} achieved via automated instrumentation of the executable. 
Our instrumentation component was coded in about $300$ lines of \texttt{Python}. In the rest 
of this section we use \program to refer to the executable to be instrumented and later attested.

Figure~\ref{fig:direct_inputs} shows the instrumentation used to implement \textbf{F3} (in MSP430 Assembly) 
which commits \program's arguments to \ilog. The instrumentation is added once: at the entry point of \program 
to log any input parameters. Lines 2-4 are already added to \program by \tinycfa to check whether \rtop is initialized 
to $OR\_MAX$. This is required by property {\bf F5} (see Section~\ref{sec:design}). Lines 5-9 are added by \acro to 
save the current stack pointer value to address $OR\_MAX$. This value determines the bottom of \program's execution 
stack and is used to detect and log data inputs. Lines 10-25 record \program's arguments (input parameters) to \ilog. 
In MSP430, function arguments are passed using up to 8 general-purposes registers $r8$--$r15$. Since the application defines 
how many arguments are passed, \acro always logs all of such registers, to guarantee that all inputs are always captured.
In this implementation, $\rtop=r4$. Hence, each register is written to the memory address pointed by $r4$. At each such write,
safety checks discussed in {\bf F5} (Section~\ref{sec:design}) are performed to assure the integrity of \ilog and \cflog in $OR$. 
Additional checks are performed to guarantee that $\rtop=r4$ never overflows the size of $OR$. Such an event is treated as a 
security violation and reported to \vrf.

Figure~\ref{fig:indirect_inputs} depicts the instrumentation used to log runtime data inputs to \ilog -- i.e., feature \textbf{F4}. 
Line 2 is a read instruction to copy contents from address pointed to by $r15$, to $r14$.
In order to define whether this is indeed a data input, at line 4, the address in $r15$ is checked against the location of the 
bottom of \program's stack, which is stored at the address of $OR\_MAX$ when \program is invoked (lines 6-9 in 
Figure~\ref{fig:direct_inputs}). Also, at line 6 in Figure~\ref{fig:indirect_inputs}, the address in $r15$ is also checked against 
the current stack pointer (always stored at register $r1$). If these checks fail, the value of the address pointed to by $r15$ 
lies outside of \program's current execution stack: it is treated as input and committed to \ilog at line 9. Otherwise, the value is 
is part of \program's current state and is not logged. Lines 10--12 check if $r4$ reached the top of $OR$, preventing overflows, 
as described in the previous paragraph.

Note that, since \acro is implemented alongside \tinycfa, it cannot be abused by control flow attacks that jump in the middle 
of the instrumented code to skip checks and/or data input logging. Such an illegal jump is itself a control-flow change, 
which is committed to \cflog by \tinycfa and thus detected by \vrf.

\lstdefinelanguage
[x64]{Assembler}     
[x86masm]{Assembler} 
%
{morekeywords={mov.b,jn,jlo,cmp.b}}

\lstset{language=[x64]{Assembler},
	basicstyle={\tiny\ttfamily},
	showstringspaces=false,
	frame=single,
	xleftmargin=2em,
	framexleftmargin=3em,
	numbers=left, 
	numberstyle=\tiny,
	commentstyle={\tiny\itshape},
	keywordstyle={\tiny\ttfamily\bfseries},
	keywordstyle=\color{purple}\tiny\ttfamily\ttfamily,
	stringstyle=\color{red}\tiny\ttfamily,
	commentstyle=\color{orange}\tiny\ttfamily,
	morecomment=[l][\color{magenta}]{\%},
	breaklines=true
}

\begin{figure}
	\begin{minipage}{0.49\linewidth}
		\begin{lstlisting}[xleftmargin=.13\textwidth, xrightmargin=.13\textwidth]
application:
; Check r4 at entry
   cmp #OR_MAX, r4
   jne .L11
	
	
	
	
	
	
	
	
	
	
	
	
	
	
	
	
	
	
	
	
   ...
		\end{lstlisting}
		\centering
		\scriptsize{(a) Before \acro instrumentation}
	\end{minipage}
	\begin{minipage}{0.49\linewidth}
		\begin{lstlisting}[xleftmargin=.13\textwidth, xrightmargin=.13\textwidth]
application:
; Check r4 at entry
   cmp #OR_MAX, r4
   jne .L11
; Save stack pointer to OR_MAX
   mov r1, @r4
   dec r4
   cmp #OR_MIN, r4
   jn .L11
; Save args registers r8 - r15
   mov r8, @r4
   dec r4
   cmp #OR_MIN, r4
   jn .L11
   mov r9, @r4
   dec r4
   cmp #OR_MIN, r4
   jn .L11
   .
   .
   .
   mov r15, @r4
   dec r4
   cmp #OR_MIN, r4
   jn .L11
		\end{lstlisting}
		\centering
		\scriptsize{(b) After \acro instrumentation}
	\end{minipage}
	\caption{Instrumentation example: Logging \program's arguments.}\label{fig:direct_inputs}
	\vspace{-0.1cm}
\end{figure}

\begin{figure}
	\begin{minipage}{0.49\linewidth}
		\begin{lstlisting}[xleftmargin=.13\textwidth, xrightmargin=.13\textwidth]
; Read from address in r15
   mov.b @r15, r14
		
		
		
		
		
		
		
		
		
		
		
		
		...
		\end{lstlisting}
		\centering
		\scriptsize{(a) Before \acro instrumentation}
	\end{minipage}
	\begin{minipage}{0.49\linewidth}
		\begin{lstlisting}[xleftmargin=.13\textwidth, xrightmargin=.13\textwidth]
; Read from address in r15
   mov.b @r15, r14
; Compare with stack range
   cmp r15, &OR_MAX
   jlo .L12
   cmp r15, r1
   jhs .L13
; Save to CF-Log if in range
.L12:
   mov @r15, @r4
   dec r4
   cmp #OR_MIN, r4
   jn .L11
.L13:
...
		\end{lstlisting}
		\centering
		\scriptsize{(b) After \acro instrumentation}
	\end{minipage}
	\caption{Instrumentation example: Logging runtime data inputs.}\label{fig:indirect_inputs}
	\vspace{-0.1cm}
\end{figure}

\section{Evaluation}\label{sec:eval}
We evaluate \acro in terms of its hardware costs and software runtime overhead 
of attested embedded operations. 

\subsection{Hardware Overhead}
Table~\ref{table:hardware_overhead} compares \acro functionality and hardware costs to prior 
runtime attestation techniques (overviewed in Section~\ref{sec:rw}). In terms of hardware, both C-FLAT~\cite{cflat} 
and OAT~\cite{oat} are based on ARM TrustZone~\cite{ARM-TrustZone} which is inapplicable to low-end MCUs.
Atrium~\cite{zeitouni2017atrium}, LO-FAT~\cite{dessouky2017fat} and LiteHAX~\cite{dessouky2018litehax} rely on 
dedicated hardware support from hash engines and branch-monitoring modules. Thus, their hardware overhead is far 
more costly than the baseline MCU (MSP430) itself. Meanwhile, \acro and \tinycfa rely exclusively on low-cost hardware 
support of the APEX's \pox architecture~\cite{apex}. Thus, they impose much lower hardware overhead, affordable even 
for such low-end MCUs. Out of all other architectures, only OAT, LiteHAX and \acro provide both \CFA and \DFA. 
Among these, \acro achieves $\approx 5 \times$ lower overhead in terms of combinatorial logic (Look-Up Tables -- LUTs) and 
$\approx 50 \times$ lower state hardware overhead (Registers) than the cheapest prior technique achieving both \CFA and \DFA,
i.e., LiteHAX.  

\begin{table}[!hbtp]
\small
	\centering
	\begin{tabular}{||P{1.4cm}||P{1.2cm}|P{1.2cm}|P{1.3cm}|P{1.3cm}||  }
		\hline
		\textbf{Technique} & \textbf{Support for \CFA} &\textbf{Support for \DFA}&\textbf{Hardware Cost -- LUTs}&\textbf{Hardware Cost -- Resigters}\\
        \hline
		MSP430 (baseline) & -- & -- & 1904 & 691 \\
		\hline
		C-FLAT & \checkmark & --    & ARM-TrustZone &  ARM-TrustZone \\
		\hline
        OAT & \checkmark & \checkmark    & ARM-TrustZone & ARM-TrustZone \\
		\hline
        Atrium & \checkmark & --    & 10640 (+559\%) & 15960 (+2308\%)\\
		\hline
		LO-FAT & \checkmark & --    & 3192 (+168\%) & 4256 (+616\%) \\
		\hline
		LiteHAX & \checkmark & \checkmark   & 1596 (+84\%) & 2128 (+308\%) \\
		\hline
		\tinycfa & \checkmark & --    & 302 (+16\%)& 44 (+6\%) \\
		\hline
        \textit{\textbf{DIALED}} & \checkmark & \checkmark    & {\bf 302 (+16\%)} & {\bf 44 $\qquad$ (+6\%) }\\
		\hline
	\end{tabular}
	\vspace*{0.2cm}
	\caption{Functionality and hardware overhead comparison of existing run-time attestation architectures}
	\label{table:hardware_overhead}
	\vspace*{-0.4cm}
\end{table}

\subsection{Experimental Analysis on Real-world Applications}\label{sec:runtime}
We evaluate \acro runtime overhead in three real-world applications. For the sake of fair comparison, we consider the exact 
same open-source applications used to evaluate \tinycfa: (1) OpenSyringePumpe~\footnote{\tiny\sf
\url{\https://github.com/naroom/OpenSyringePump/blob/master/syringePump/syringePump.ino}} -- a medical syringe pump; (2) 
FireSensor~\footnote{\tiny\sf \url{https://github.com/Seeed-Studio/LaunchPad_Kit/tree/master/Grove_Modules/temp_humi_sensor}}; 
and (3) UltrasonicRanger~\footnote{\tiny\sf \url{https://github.com/Seeed-Studio/LaunchPad_Kit/tree/master/Grove_Modules/ultrasonic_ranger}} -- 
a sensor used in vehicles to measure distance from obstacles.

\begin{figure*}[hbtp]
	\centering
	\subfigure[Total code size]
	{\includegraphics[width=0.66\columnwidth]{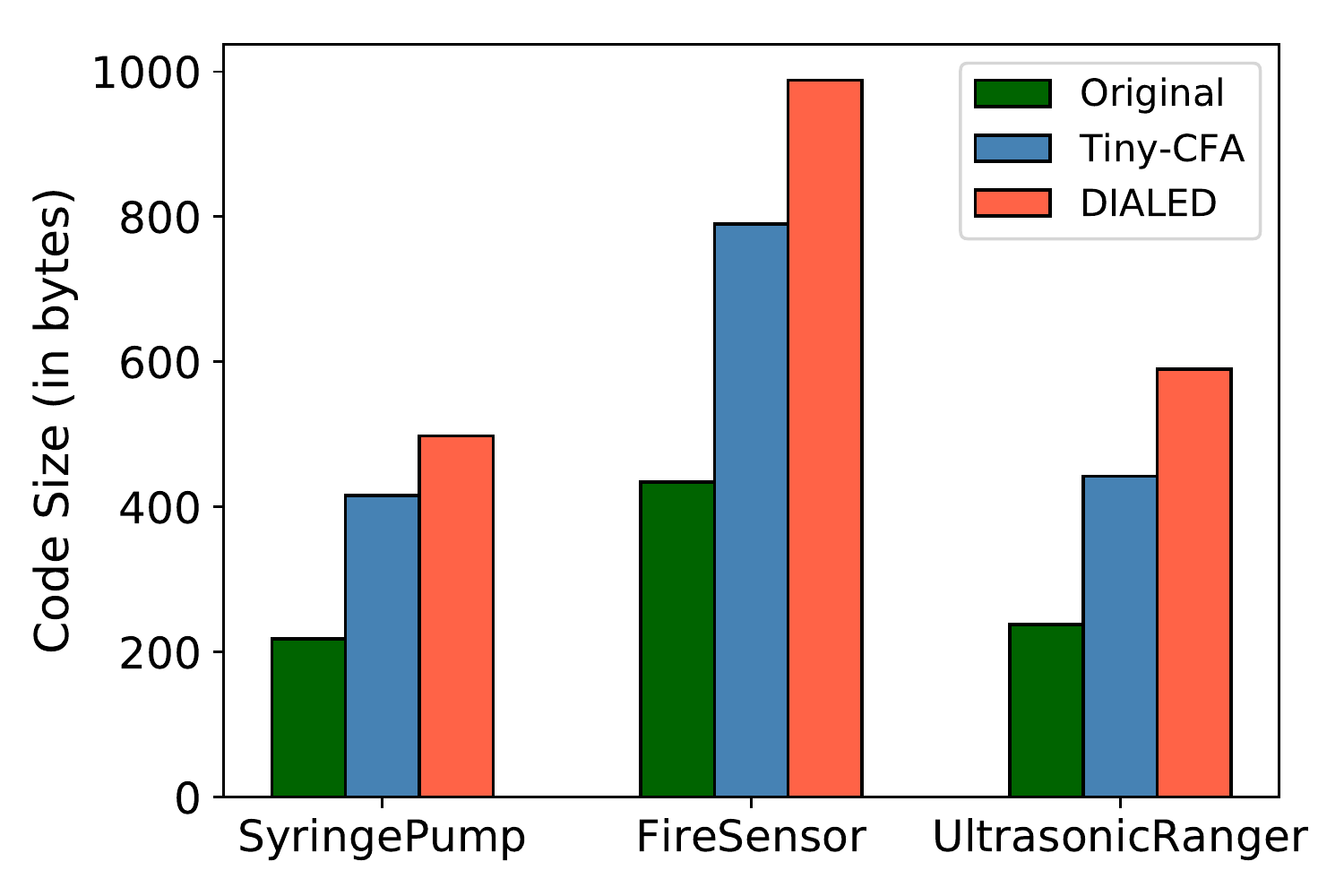}\label{fig:codesize_graph}}
	\subfigure[Runtime]
	{\includegraphics[width=0.66\columnwidth]{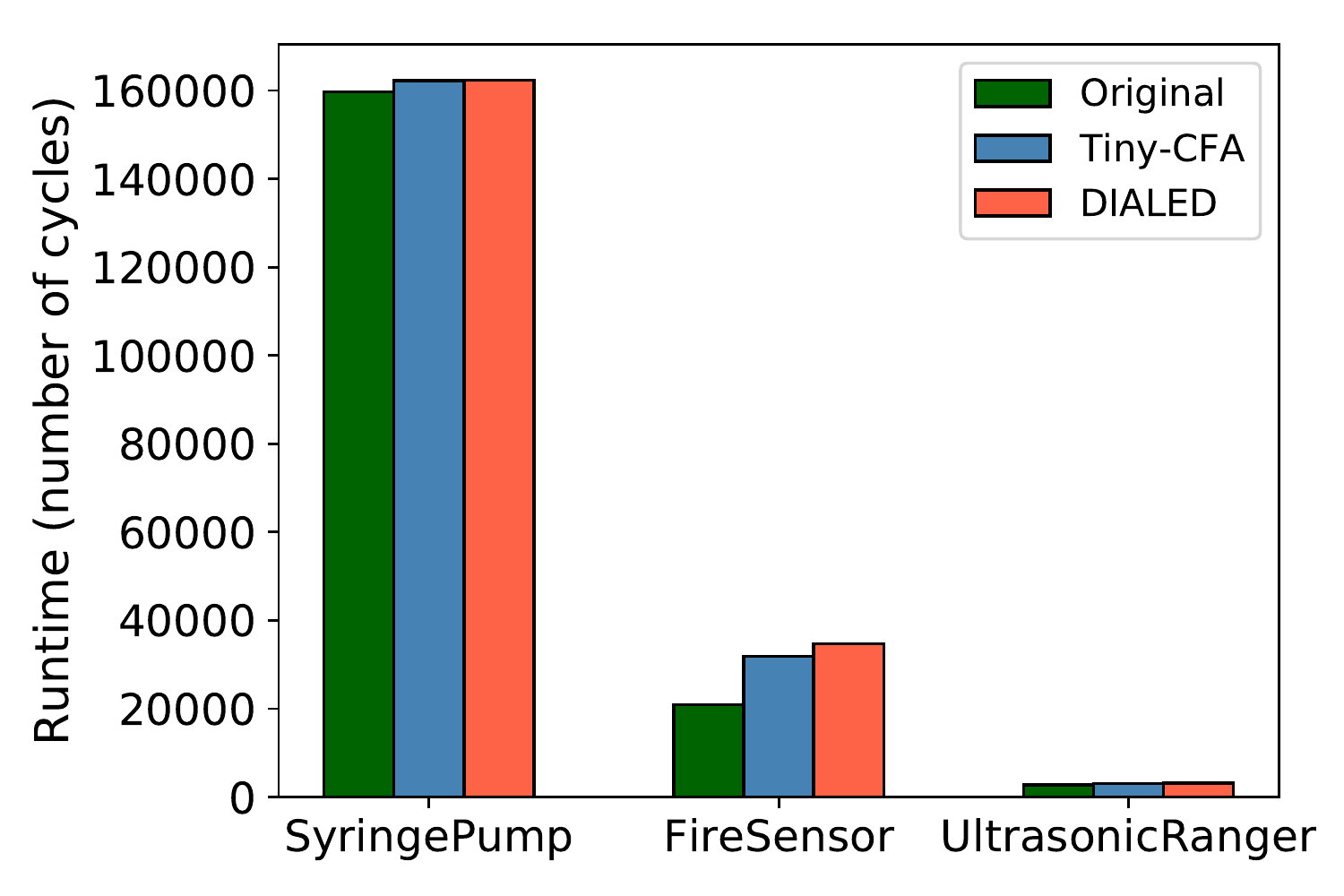}\label{fig:runtime_graph}}
	\subfigure[Log size]
	{\includegraphics[width=0.66\columnwidth]{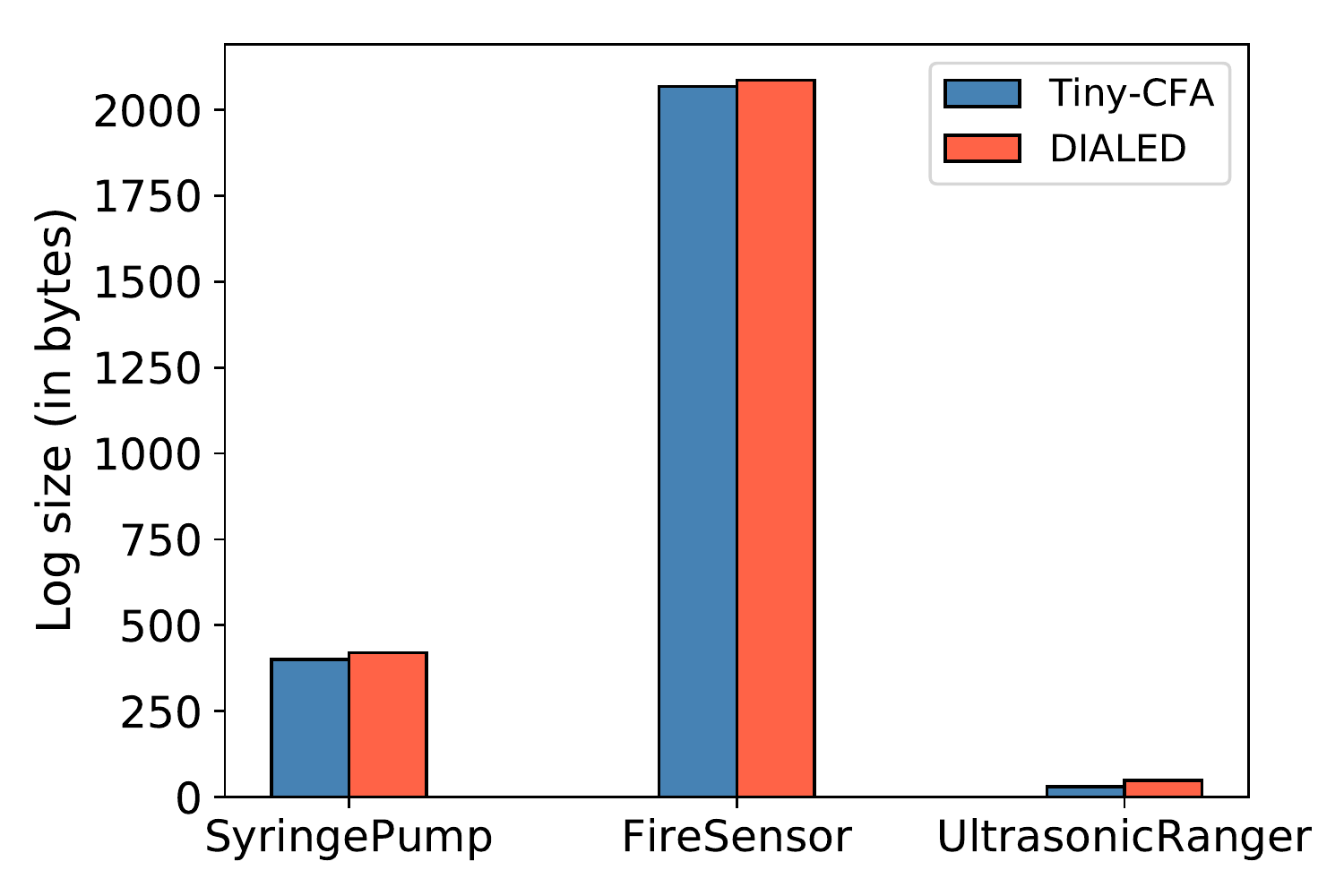}\label{fig:log_sizes}}
	\caption{Comparison of embedded operation runtime costs: unmodified operation, \tinycfa (\CFA only), and \acro (\CFA+\DFA).}\label{fig:comparison}
\end{figure*}

We consider all sources of runtime overhead imposed by code instrumentation in these techniques: code size increase, runtime 
(CPU cycles), and the size of the attestation log inside $OR$, including \ilog and \cflog. Figures~\ref{fig:codesize_graph} 
and~\ref{fig:runtime_graph} compare results unmodified applications, with the same applications instrumented by \tinycfa 
(\CFA guarantee only), and the same applications instrumented by \acro (both \CFA and \DFA guarantees). As these results 
demonstrate, the overhead in both cases is dominated by the instrumentation required for \CFA.
On top of \tinycfa, \acro code size and runtime increases range between $1\%-20\%$. This is due to additional 
instructions introduced by \acro instrumentation, as described in Section~\ref{sec:implementation}.
Figure~\ref{fig:log_sizes} shows the total $OR$ size required to store the execution information. Recall that
\acro requires storage of both \ilog and \cflog to enable detection of both control-flow and data-only attacks.
Size requirement for these logs vary widely depending on the type of application (control-flow- or data-input-intensive). 
In general, we observe a small increase in $OR$ size. This is due to the data input definition from Section~\ref{sec:design}, 
which allows \acro to only log relevant data inputs while retaining all necessary information for \vrf's abstract execution of \prv's embedded operation.

\textit{\textbf{Remark}:} we do not compare runtime overhead of \acro with \DFA architectures LiteHAX and OAT, since 
these techniques rely on specific hardware support implemented in different CPU architectures (generally higher-end platforms), 
with different applications. 

In summary, even though \acro's overhead is certainly not negligible, it is well within the capabilities of low-end MCUs and 
suitable for practical purposes. Specifically, instrumented binary sizes are within the MCUs memory budget, its 
runtime is reasonable, and log sizes are small enough to fit into data memory without encroaching on the stack.
We believe this to be a reasonable price for the benefit of detecting any runtime compromise in low-end MCUs.

\section{Related Work}\label{sec:rw}
\noindent\textbf{Control-Flow and Data-Flow Integrity}~\cite{abadi2009control,castro2006securing} are techniques
for prevention or detection of data and control-flow corruptions in real-time, locally at \prv, in contrast with after-the-fact 
detection by \vrf. All such techniques depend on complex software support (e.g., high-end operating systems) and/or 
heavyweight address-based and value-based integrity checks at runtime. Unfortunately, all these options are inapplicable
to simple MCUs.

\noindent\textbf{(Static) Remote Attestation (\RA)}~\cite{smart,vrasedp,simple,hydra,tytan,trustlite} is used by \vrf to check 
if a remote \prv has the proper software. By itself, static \RA does not provide any runtime guarantees. However, it is used 
as a building block for most runtime attestation techniques, including \CFA and \DFA. \acro itself is build atop APEX \pox 
functionality. In turn, APEX relies on VRASED~\cite{vrasedp} -- a formally verified static \RA architecture -- 
to implement \pox.

\noindent\textbf{Runtime Attestation} includes techniques such as \CFA and \DFA. C-FLAT~\cite{cflat} is the earliest 
\CFA architecture that uses ARM TrustZone Secure World~\cite{trustzone} to implement \CFA, by instrumenting the 
executable with context switches between TrustZone Normal and Secure worlds. At each control flow altering instruction, 
execution is trapped into Secure World and the control flow path is logged to protected memory. To remove TrustZone dependence, 
LO-FAT~\cite{dessouky2017fat} and LiteHAX~\cite{dessouky2018litehax} implement CFA using stand-alone hardware modules: 
a branch monitor and a hash engine. Atrium~\cite{zeitouni2017atrium} enhances aforementioned CFA techniques by securing 
them against physical adversaries that intercept instructions as they are fetched to the CPU. Though less expensive than C-FLAT, 
such hardware components are still not affordable for low-end MCUs, since their cost (in terms of price, size, and energy consumption) 
is higher than that of a low-end MCU itself. OAT~\cite{oat} and LiteHax~\cite{dessouky2018litehax} also provide 
\DFA (in addition to \CFA). However, similar to aforementioned techniques, they are too costly for low-end MCUs.

\section{Conclusions}\label{sec:conclusion}
We design and implement \acro, the first Data-Flow Attestation (\DFA) approach targeting lowest-end MCUs. 
\acro is composed with \tinycfa, a Control-Flow Attestation (\CFA) architecture, thus enabling detection of both control-flow 
and data-flow attacks at runtime. We discuss \acro's security and evaluate its performance on real embedded applications, showing that \acro's overhead is well within the capabilities of some of the most resource-constrained MCUs.
\\\\
\noindent {\bf Acknowledgments:} We thank DAC'21 anonymous referees for their helpful comments.
This research was supported in part by funding from Army Research Office (ARO) contract
W911NF-16-1-0536, Semiconductor Research Corporation (SRC) contract 2019-TS-2907, as well as 
NSF Awards 1956393 (SATC) and 1840197 (CICI).

\bstctlcite{IEEEexample:BSTcontrol}
\bibliographystyle{IEEEtranS}

{\small
	\linespread{0.90}
	\bibliography{IEEEabrv,references}
}
\end{document}